\documentstyle[12pt,aps,prl,epsf]{revtex}
\draft
\begin{document}

\author{D. L. Zhou, G. R. Jin and C. P. Sun$^{a,b}$}
\address{Institute of Theoretical Physics, Academia Sinica, \\
P. O. Box 2735, Beijing 100080, China}
\title{Second Quantization of Cini Model for High Order Quantum Decoherence in
Quantum Measurement}
\maketitle

\begin{abstract}
By making the second quantization for the Cini Model of quantum
measurement without wave function collapse [M. Cini, Nuovo
Cimento, B73 27(1983)], the second order quantum decoherence
(SOQD) is studied with a two mode boson system interacting with an
idealized apparatus composed by two quantum oscillators. In the
classical limit that the apparatus is prepared in a Fock state
with a very large quantum number, or in a coherent state with
average quantum numbers large enough, the SOQD phenomenon appears
similar to the first order case of quantum decoherence.
\end{abstract}
 \draft

%@twocolumnfalse\endcsname
\pagenumbering{roma}

\thispagestyle{empty} \vspace{15mm} \widetext

\widetext
\vspace{4mm}

%\vskip2pc]
\pagenumbering{arabic} %\begin{multicols}{2}

\section{Introduction}

In usual the quantum coherence is  reflected by the spatial
interference of two or more ``paths'' in terms of single particle
wave function. Correspondingly the decoherence phenomenon losing
coherence can be understood in term of an ``which-path'' detection
implied by the quantum entanglement of the considered system with
the environment or the measuring apparatus [1-3]. Most recently,
we have shown [4,5] that this more simple, but most profound
observation can be also implemented in the many particle picture
to account for the losing of the high order quantum coherence
(HOQC) described by the high order correlation function [6,7]. In
this letter we will give a detailed study of this novel context
specified for the quantum measurement problem.

To this end we first make the second quantization for the Cini
Model of quantum measurement without wave function collapse [8] to
obtain a modelled system --- a two mode boson system interacting
with an idealized measuring apparatus composed also by two quantum
oscillators. Then, the second order quantum decoherence (SOQD) is
studied with this model in the classical limit that the apparatus
is prepared in a Fock state with a very large quantum number, or
in a coherent state with average quantum numbers large enough.

The crucial point that we understand the higher order quantum
decoherence problem in the ``which-path'' picture is to introduce
the concept of the multi-particle wave amplitude (MPWA), whose
norm square is just the high order correlation function [4,5].
Before the measurement, as an effective wave function, this
multi-time amplitude can be shown to be a supposition of several
components. When the an apparatus entangles with them to make an
effective measurement, the high order quantum coherence loses
dynamically. This decoherence process can be explained as a
generalized ``which-path'' measurement for the defined
multi-particle paths in the MPWA .

\section{Second Quantization of Cini Model }

The original Cini model for quantum measurement emphasizes the
production of quantum entanglement between the states of measured
system S (a two-level system) and the measuring apparatus D ---
many indistinguishable particles with two possible modes $\omega
_1$ and $\omega _2$. The two states $u_g$ and $u_e $ of S have
different interaction strengths $d_g$ and $d_e$ with D. Then,
the large number N of ``ionized'' particles in the ``ionized'' mode $%
\omega _2$ transiting from the ``un-ionized'' model $\omega _2$
shows this quantum entanglement. In the following we wish to make
a second quantization for the system components to built a novel
model for SOQD\ with Hamiltonian

\begin{eqnarray}
\hat{H}_0 &=&\omega _e\hat{b}_e^{\dagger }\hat{b}_e, \\
\hat{V} &=&\omega _1\hat{a}_1^{\dagger }\hat{a}_1+\omega _2\hat{a}%
_2^{\dagger }\hat{a}_2+(d_e\hat{b}_e^{\dagger }\hat{b}_e+d_g\hat{b}%
_g^{\dagger }\hat{b}_g)(\hat{a}_1^{\dagger }\hat{a}_2+\hat{a}_2^{\dagger }%
\hat{a}_1),
\end{eqnarray}
where $\hat{H}_0$ is the free Hamiltonian of the system, $\hat{V}$ the free
Hamiltonian of the apparatus D plus the interaction between S and D; and $%
\hat{b}_e^{\dagger }(\hat{b}_e), \hat{b}_g^{\dagger }(\hat{b}_g)$
the creation (annihilation) operators of two modes labelled by
index $e$ and $g$. Their frequencies are $\omega _e$ and $\omega
_g=0$ respectively. The operators $\hat{a}_j^{\dagger
}(\hat{a}_j)$ are creation (annihilation) operators of the modes
which labelled by index $j$ for the mode frequency $\omega _j$,
${j=1,2}$.
The frequency-dependent constant $%
d_e$ ($d_g$) measures the coupling constant between the $e$ ($g$)
mode of the system and the apparatus. The most important feature
of the model is that $[H_0,V]=0$, i.e. the system does not
dissipate energy to the apparatus. Notice this model is equivalent
to the generalized Cini model with many-levels given by us [9].

Starting with this concrete model we first consider the meaning of
the ``path'' for  the high order quantum correlation in the free
particle case. The typical example of the higher order quantum
coherence is that the single-component state $|1_e,1_g\rangle $ of
the two independent particles shows its quantum coherence in its
second order quantum correlation function $G^{(2)}(t_1,t_2)$,
which can just be written as the norm square $G^{(2)}=|\psi |^2$
[10] of the equivalent ``two-time wave function'' $\Psi (t_1,t_2)$
(it was also called the bipparticle wavepacket [11] for photons),
namely,
\begin{eqnarray}
G^{(2)} &=&\langle 1_e1_g|\hat{\phi}^{\dagger }(t_1)\hat{\phi}%
^{\dagger }(t_2)\hat{\phi}(t_2)\hat{\phi}(t_1)|1_e,1_g\rangle   \nonumber \\
\mbox{} &=&|\langle 0,0|\hat{\phi}(t_2)\hat{\phi}(t_1)|1_e,1_g\rangle
|^2\equiv |\Psi (t_1,t_2)|^2
\end{eqnarray}
Here, we define a ``measuring'' field operator of two modes $g$
and $e$
\begin{equation}
\hat{\phi}=c_g\hat{b}_ge^{-i\omega _gt}+c_e\hat{b}_ee^{-i\omega
_et}\equiv c_g(t)\hat{b}_e+c_e(t)\hat{b}_e.
\end{equation}
The two time wave function $\Psi (t_1,t_2)$ can be understood in
terms of the two ``paths'' picture from the initial state
$|1_g,1_e\rangle $ to the finial one $|0,0\rangle $ [5]: \vskip
5mm

\begin{center}
\begin{tabular}{|c|}
\hline $
\begin{array}{ccccc}
|1_g,1_e\rangle & \stackrel{c_e(t_1)}{\longrightarrow } &
|1_g,0_e\rangle &
\stackrel{c_{_g}(t_2)}{\longrightarrow } & |0,0\rangle \\
& \searrow \stackrel{c_g(t_1)}{} &  & \stackrel{c_e(t_2)}{}\nearrow &  \\
&  & |1_e,0_g\rangle &  &
\end{array}
$ \\ \hline
\end{tabular}
\end{center}
%\begin{center}
%\begin{figure}[tbp]
%\epsfxsize=16.8cm \epsffile{whichway.eps}
%\end{figure}
%\end{center}

Obviously, they are just associated with the two amplitudes forming a
coherent superposition

\begin{eqnarray}
\Psi (t_1,t_2) &=&\langle 0,0|\hat{\phi}(t_2)\hat{\phi}(t_1)|1_e,1_g\rangle
\\
&=&c_ec_ge^{-i\omega _et_2-i\omega _gt_1}+c_gc_ee^{-i\omega _{_g}t_2-i\omega
_{_e}t_1}
\end{eqnarray}
Correspondingly, the second order correlation function
\begin{equation}
G^{(2)}=2|c_ec_g|^2[1+\cos ([\omega _g-\omega _e][t_2-t_1])]
\end{equation}
shows the HOQC in the time domain. The above observation for the
second order quantum coherence can also be discovered in the
higher order case. It is noticed that our present arguments  will
be based on the equivalent field operator $\hat{\Phi}=\sum
c_n\hat{b}_n$ is specified for a quantum measurement to a
superposition state $|\phi \rangle =\sum c_n|n\rangle$. In
reference [5], we have point out its observability in an idealized
cavity QED experiment.

\section{Many-Particle ``Which-Way'' Detection}

In the case with interaction, we consider the generalized second
order correlation functions
\begin{equation}
G[t,t^{\prime },\hat{\rho}(0)]=Tr(\hat{\rho}(0)\hat{B}^{\dagger }(t)\hat{B}%
^{\dagger }(t^{\prime })\hat{B}(t^{\prime })\hat{B}(t)),
\end{equation}
with respect to ``measuring'' field operator [4]. It is defined as
a functional of the density operator $\hat{\rho}(0)$ of the whole
system for a given time $0$. Here, the bosonic field operator
\begin{eqnarray}
\hat{B}(t) &=&\exp (i\hat{H}t)[c_1\hat{b}_g+c_2\hat{b}_e]\exp (-i\hat{H}t) \\
&=&c_1\exp (i\hat{V}t)[c_1\hat{b}_g+\hat{b}_ec_2\exp (-i\omega _et)]\exp (-i%
\hat{V}t)
\end{eqnarray}
describes a specific quantum measurement with respect to the superpositions $%
|+\rangle =$ $c_1|e\rangle $ $+c_2|g\rangle $ and $|-\rangle =$ $%
c_2|e\rangle $ $-c_1|g\rangle $ where $c_1$ and $c_2$ satisfy the
normalization relation $|c_1|^2+|c_2|^2=1$. Without loss of the generality,
we take $c_1=c_2=1/\sqrt{2}$ standing for a measurement as follows.

To examine whether the classical feature of the apparatus causes
the second order decoherence or not, we consider the whole system
in an initial state
\begin{equation}
|\psi (0)\rangle =|1_g,1_e\rangle \otimes |\phi (0)\rangle ,
\end{equation}
where $|\phi (0)\rangle $ is the initial state of the apparatus.
In the case with interaction, in stead of defining the equivalent
``two-time wave function'' in the case of free particle, we define
an effective two-time state vector
\begin{equation}
|\psi _B(t,t^{\prime })\rangle =\hat{B}(t^{\prime })\hat{B}(t)|\psi
(0)\rangle .
\end{equation}
to re-write the second order correlation function as
\begin{equation}
G[t,t^{\prime },\hat{\rho}(0)]=\langle \psi _B(t,t^{\prime })|\psi
_B(t,t^{\prime })\rangle
\end{equation}
It is interested that the effective state vector can be evaluated as the
superposition
\begin{eqnarray}
|\psi _B(t,t^{\prime })\rangle  &=&\frac 12e^{i\hat{V}(0,0)t^{\prime }}[\exp
(-i\omega _et^{\prime })e^{-i\hat{V}(1,0)t^{\prime }}e^{i\hat{V}(1,0)t}+ \\
&&\exp (-i\omega _et)e^{i\hat{V}(0,0)t^{\prime }}e^{-i\hat{V}(0,1)t^{\prime
}}e^{i\hat{V}(0,1)t}]e^{-i\hat{V}(1,1)t}|\{0_j\}\rangle \otimes
|0_g,0_e\rangle
\end{eqnarray}
of two components with respect to the two paths from the initial two particle state $%
|1_g,1_e\rangle $ to the two particle vacuum $|0_g,0_e\rangle $ .
It should be noticed that the effective actions of the apparatus
\begin{equation}
\hat{V}(m,n)\equiv \sum_j\hat{V}_j(m,n)=\sum_j\omega _j\hat{a}_j^{\dagger }%
\hat{a}_j+\sum_j(d_e(\omega _j)m+d_g(\omega _j)n)(\hat{a}_j^{\dagger }+\hat{a%
}_j)
\end{equation}
can label the different paths and thus lead to the higher order
quantum decoherence. The above result clearly demonstrates that,
in presence of the apparatus, the different probability amplitudes
($\sim $ $\exp (-i\omega
_gt^{\prime })$ and $\exp (-i\omega _et)$) from $|1_g,1_e\rangle $ to $%
|0_g,0_e\rangle $ entangle with the different states ($\frac 12e^{i\hat{V}%
(0,0)t^{\prime }}e^{-i\hat{V}(1,0)t^{\prime }}e^{i\hat{V}(1,0)t}e^{-i\hat{V}%
(1,1)t}|\{0_j\}\rangle $ and $\frac 12e^{i\hat{V}(0,0)t^{\prime }}e^{-i\hat{V%
}(0,1)t^{\prime
}}e^{i\hat{V}(0,1)t}e^{-i\hat{V}(1,1)t}|\{0_j\}\rangle $ ) of the
apparatus. This is just physical source of the higher order
quantum decoherence.

In the following calculation, the second order correlation function
\begin{equation}
G[t,t^{\prime },\hat{\rho}(0)]=\frac 12+\frac{e^{i\omega _e(t-t^{\prime })}}%
4F+\frac{e^{-i\omega _e(t-t^{\prime })}}4F^{*},
\end{equation}
is expressed explicitly in terms of the decoherence factor
\begin{equation}
F=\langle \phi (0)|e^{i\hat{V}(1,1)t}e^{-i\hat{V}(0,1)t}e^{i\hat{V}%
(0,1)t^{\prime }}e^{-i\hat{V}(1,0)t^{\prime }}e^{i\hat{V}(1,0)t}e^{-i\hat{V}%
(1,1)t}|\phi (0)\rangle
\end{equation}
which determines the extent of coherence or decoherence in the
second order case.

\section{Dynamic High-Order Quantum Decoherence}

In the following, to given the factor $F$ explicitly, the normal
ordering technique [12] is adopted to calculate the second order
decoherence factor $F$. The calculation is carried out  in six
steps.

At the $k$-th step the evolution is dominated by the
step-Hamiltonian
\[
\hat{h}^k={\alpha _1^k}\hat{a}_1^{\dagger }\hat{a}_1+{\alpha _2^k}\hat{a}%
_2^{\dagger }\hat{a}_2+{\beta ^k}(\hat{a}_1^{\dagger }\hat{a}_2+\hat{a}%
_2^{\dagger }\hat{a}_1),{k=1,2,\cdots ,6}
\]
during the time period $t_k$. The coefficients $\{\alpha
_1^k,\alpha _2^k,\beta ^k,t_k\}$ take different values in
different steps:
\begin{eqnarray}
&&\alpha _1^1=\omega _1,\alpha _2^1=\omega _2,\beta ^1=d_e+d_g,t_1=t,
\nonumber \\
&&\alpha _1^2=-\omega _1,\alpha _2^2=-\omega _2,\beta ^2=-d_e,t_2=t,
\nonumber \\
&&\alpha _1^3=\omega _1,\alpha _2^3=\omega _2,\beta ^3=d_e,t_3=t^{\prime },
\nonumber \\
&&\alpha _1^4=-\omega _1,\alpha _2^4=-\omega _2,\beta ^4=-d_g,t_4=t^{\prime
},  \nonumber \\
&&\alpha _1^5=\omega _1,\alpha _2^5=\omega _2,\beta ^5=d_g,t_5=t,  \nonumber
\\
&&\alpha _1^6=-\omega _1,\alpha _2^6=-\omega _2,\beta ^6=-d_e-d_g,t_6=t.
\end{eqnarray}
Assume that at the k-th step the evolution operator $\hat{u}^k(t)$ can be
written in the normal order as
\begin{equation}
\hat{u}^k(t)=\aleph \{e^{{A^k}(t)\hat{a}_1^{\dagger }\hat{a}_1+{B^k}(t)\hat{a%
}_2^{\dagger }\hat{a}_2+{C^k}(t)\hat{a}_1^{\dagger }\hat{a}_2+{D^k}(t)\hat{a}%
_2^{\dagger }\hat{a}_1}\}.
\end{equation}
The advantage of this form is that, when calculating the average
of the operator in the coherent state $|\alpha ,\beta \rangle $,
we only need to replace the annihilation operators with the
corresponding complex values. This evolution operators satisfy the
$Schr\ddot{o}dinger$ equation $i\frac
d{dt}\hat{u}^k=\hat{h}^k\hat{u}^k$. To take the expectation values
of the above equations in the coherent state $|\alpha ,\beta
\rangle $, the coefficients satisfy the following system of
equations:
\begin{eqnarray}
i\frac{dA^k}{dt} &=&\alpha _1^k(A^k+1)+\beta ^kD^k,  \nonumber \\
i\frac{dB^k}{dt} &=&\alpha _2^k(B^k+1)+\beta ^kC^k,  \nonumber \\
i\frac{dC^k}{dt} &=&\alpha _1^kC^k+\beta ^k(B^k+1),  \nonumber \\
i\frac{dD^k}{dt} &=&\alpha _2^kD^k+\beta ^k(A^k+1).
\end{eqnarray}
The solution of the system of equations is
\begin{eqnarray}
A^k+1 &=&e^{-i(\alpha _1^k+\alpha _2^k)t_k/2}(\cos (\Gamma ^kt_k)+\frac{%
i(\alpha _2^k-\alpha _1^k)}{2\Gamma ^k}\sin (\Gamma ^kt_k)),  \nonumber \\
B^k+1 &=&e^{-i(\alpha _1^k+\alpha _2^k)t_k/2}(\cos (\Gamma ^kt_k)-\frac{%
i(\alpha _2^k-\alpha _1^k)}{2\Gamma ^k}\sin (\Gamma ^kt_k)),  \nonumber \\
C^k=D^k &=&-\frac{i\beta ^k}{\Gamma ^k} e^{-i(\alpha _1^k+\alpha
_2^k)t_k/2}\sin (\Gamma ^kt_k),
\end{eqnarray}
where
\[
\Gamma ^k=\sqrt{(\frac{\alpha _2^k-\alpha _1^k}2)^2+{\beta ^k}^2}
\]
Notice that the above results have been before given in ref.[12],
but the original ones contain some minor misprints. Here, we have
corrected them. Then we obtain
\begin{eqnarray}
&&e^{-i\hat{h}^kt_k}|\alpha ^{k-1},\beta ^{k-1}\rangle   \nonumber \\
&=&|(A^k+1)\alpha ^{k-1}+C^k\beta ^{k-1},(B^k+1)\beta
^{k-1}+D^k\alpha ^{k-1}\rangle
\nonumber \\
&\equiv &|\alpha ^k,\beta ^k\rangle.
\end{eqnarray}
From the above equation, it is obvious that, when the apparatus is
initially in the product coherent state $|\alpha ^0,\beta
^0\rangle $, after six steps of evolution, the final state of the
apparatus remains in a product coherent state $|\alpha ^6(\alpha
^0,\beta ^0),\beta ^6(\alpha ^0,\beta ^0)\rangle $.

To consider the classical feature of the apparatus, two specific
initial states will be studied with different classical
correspondences. And we give the numerical results respectively
thereafter. The first case is that the initial state of the
apparatus takes $|\phi (0)\rangle =|0,\beta \rangle $. When the
norm of $\beta $ goes to infinity, it corresponds to classical
field in some sense. In this case, we can obtain the decoherence
factor and therefore the second order correlation function becomes
\begin{equation}
G[t,t^{\prime },\hat{\rho}(0)]=\frac 12+(\frac{e^{i\omega _e(t-t^{\prime })}}%
4\langle 0,\beta |\alpha ^6(0,\beta ),\beta ^6(0,\beta )\rangle +h.c.).
\end{equation}
A typical case of the numerical result of the above equation is
given in FIG.$1$.
\begin{figure}[tbsp]
\begin{tabbing}
\= \epsfxsize=8.3cm\epsffile{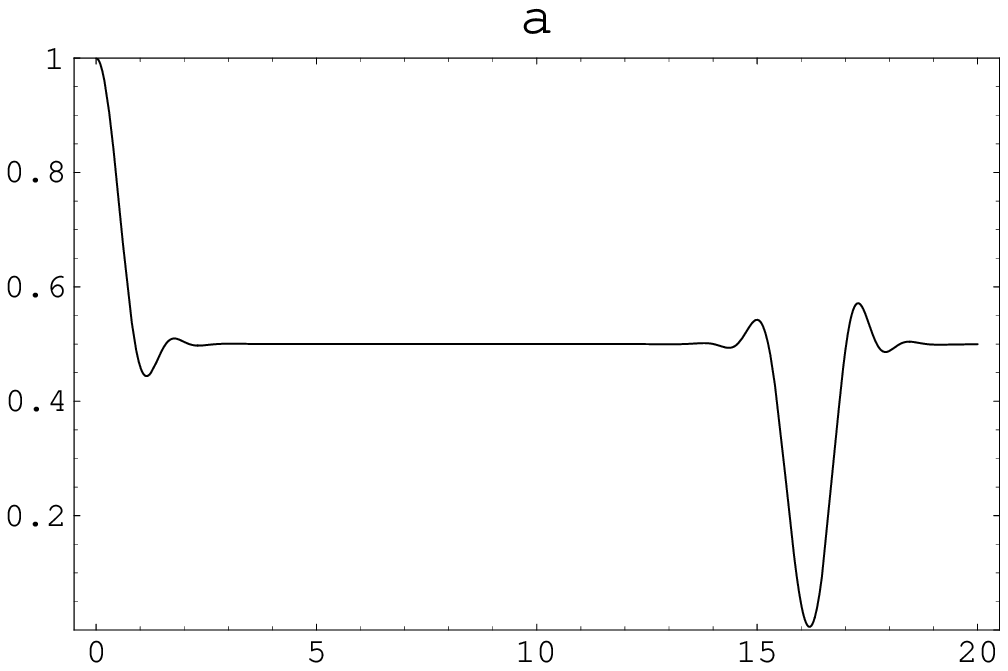} \=
\epsfxsize=8.3cm\epsffile{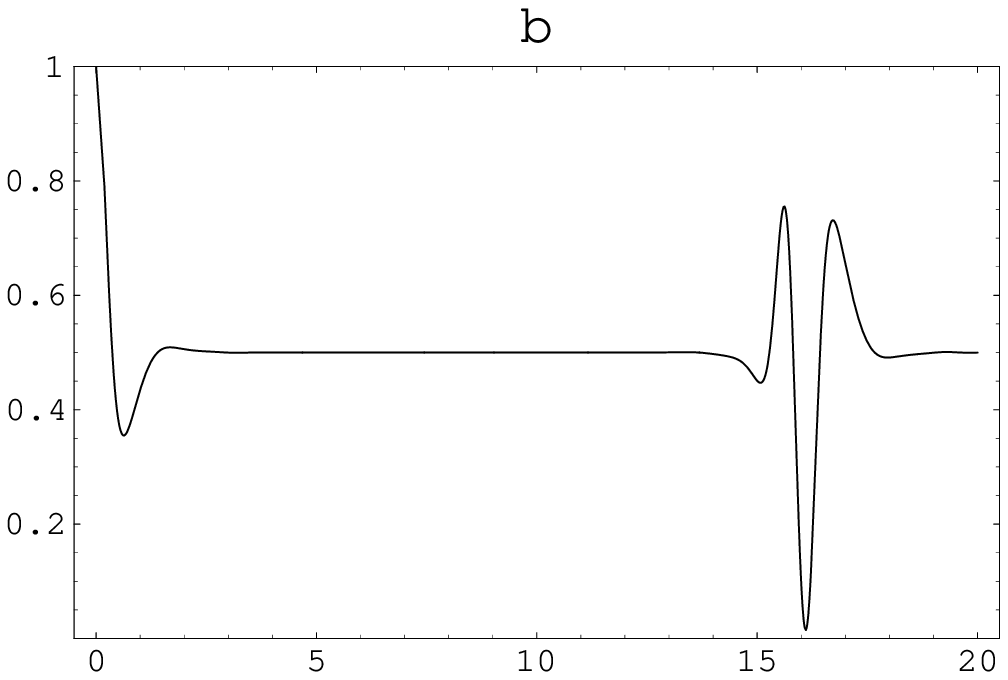}\\ \>
\epsfxsize=8.3cm\epsffile{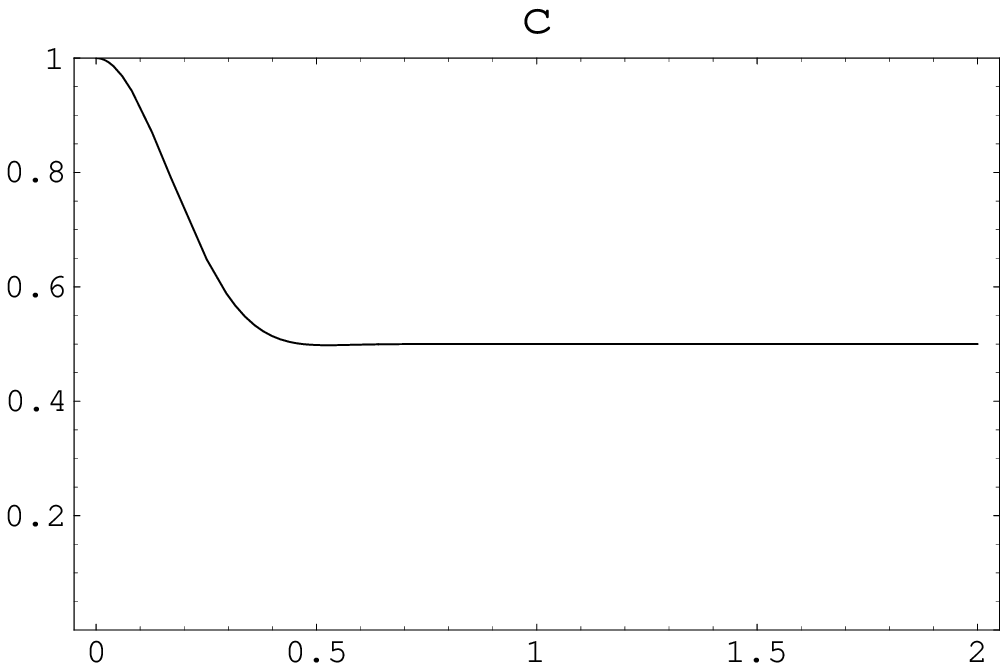} \>
\epsfxsize=8.3cm\epsffile{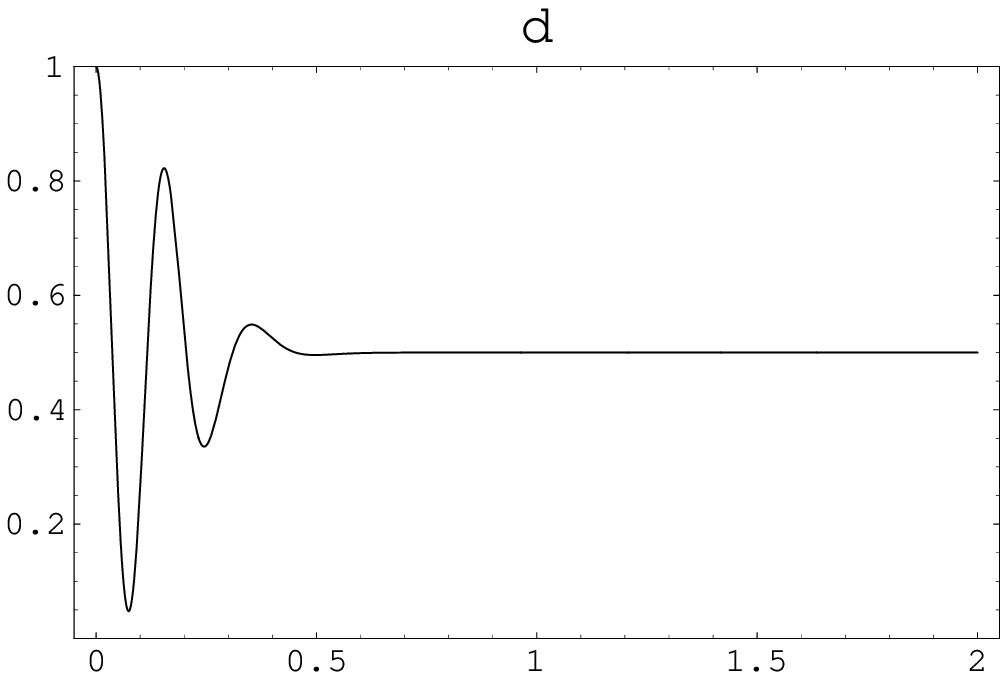}\\ \>
\epsfxsize=8.3cm\epsffile{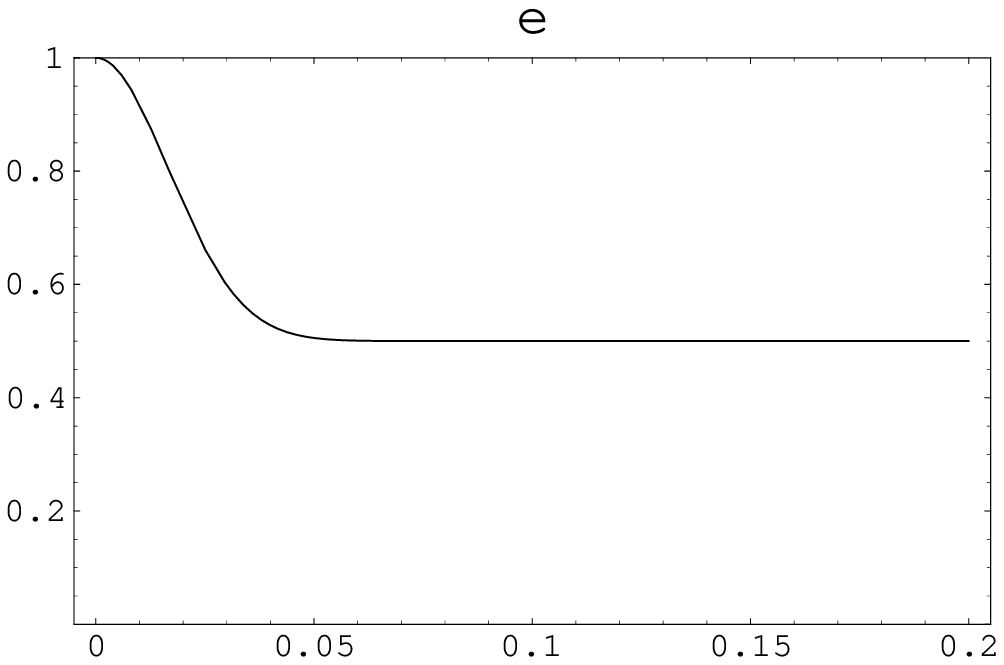} \>
\epsfxsize=8.3cm\epsffile{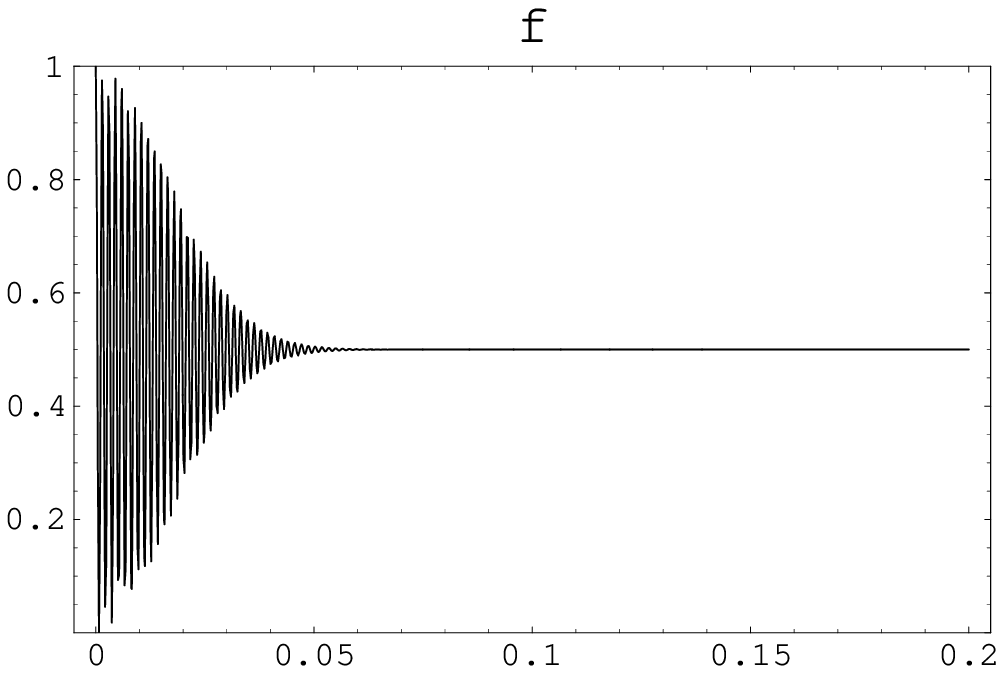}
\end{tabbing}
\caption{The horizontal axe denotes time period $t^{\prime }-t$, the
vertical axe denotes the second order correlation function $G[t,t^{\prime },%
\hat{\rho}(0)]$, parameters ${\omega _1=0.2,\omega
_2=1.3,d_e=0.8,d_e=0.2,\omega _e=1.0}$, (a)${N=10,t=0}$, (b)${N=10,t=10}$,
(c)${N=10^2,t=0}$, (d)${N=10^2,t=10}$, (e)${N=10^4,t=0}$, (f)${N=10^4,t=10}$%
. }
\end{figure}

In FIG.$1$, we observe that the second order correlation function
is an explicit function of both the time interval $t^{\prime }-t$
and the time $t$. With the increasing of time $t$, it obviously
oscillate faster and faster. It is all observed that, as the
average particle number of the coherent state increases, the
second order correlation function decoheres in a shorter time
scale. The decoherence rate is independent of the time $t$. The
later observation implies that, when the average particle number
approaches to infinity, the second order correlation function will
decohere in very short time, and the quantum revivals can not be
observed in a finite time period.

The second case is that the initial state of the apparatus takes a
Fock number state $|\phi (0)\rangle =|0,N\rangle $. When the
number $N$ approaches into infinity, it also corresponds to
classical field in some sense. In this case, we can obtain the
decoherence factor and therefore the second order correlation
function
\begin{equation}
G[t,t^{\prime },\hat{\rho}(0)]=\frac 12+(\frac{e^{i\omega _e(t-t^{\prime })}}%
4\int \frac{d^2\beta }\pi \langle 0,N|\alpha ^6(0,\beta ),\beta ^6(0,\beta
)\rangle \langle \beta |N\rangle +h.c.).
\end{equation}
A typical numerical result of the above equation with the same
parameters as in FIG. 1. is given in FIG.2.

\begin{figure}[tbp]
\begin{tabbing}
\= \epsfxsize=8.3cm\epsffile{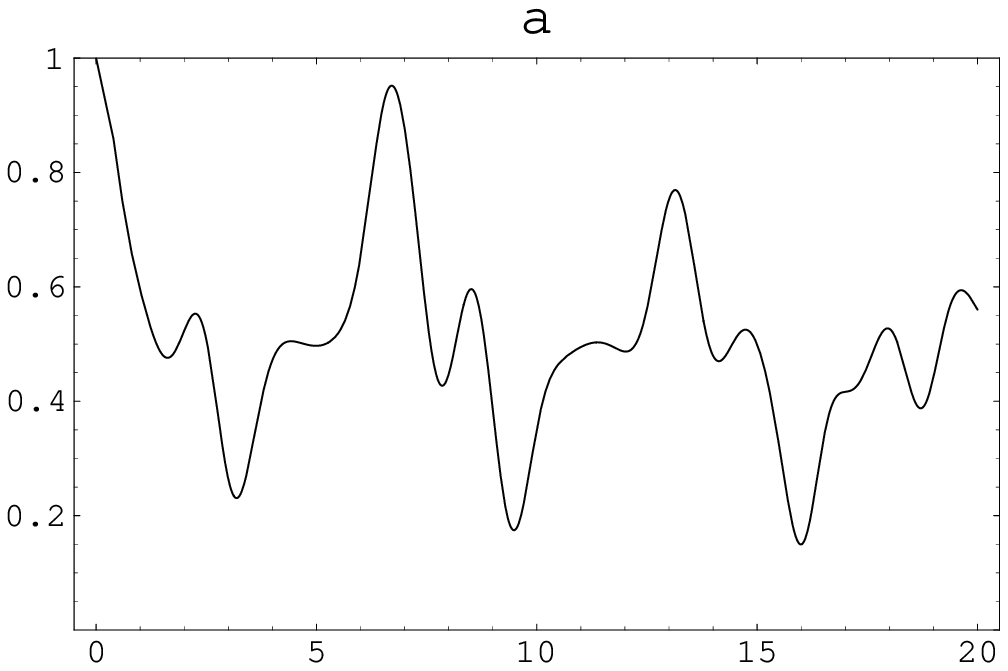}
\= \epsfxsize=8.3cm\epsffile{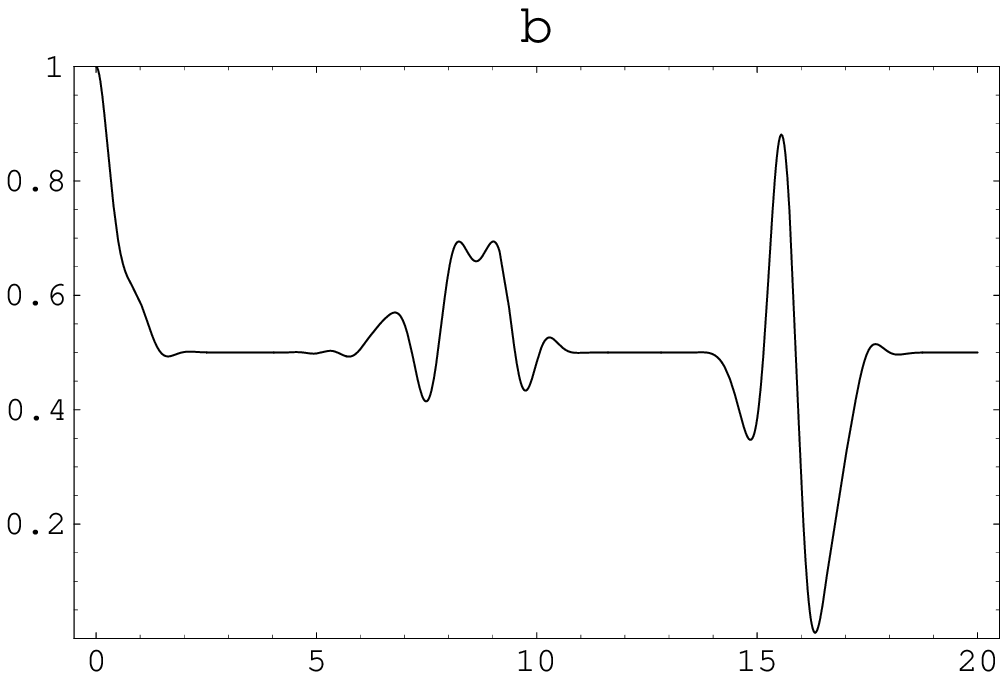}\\
\> \epsfxsize=8.3cm\epsffile{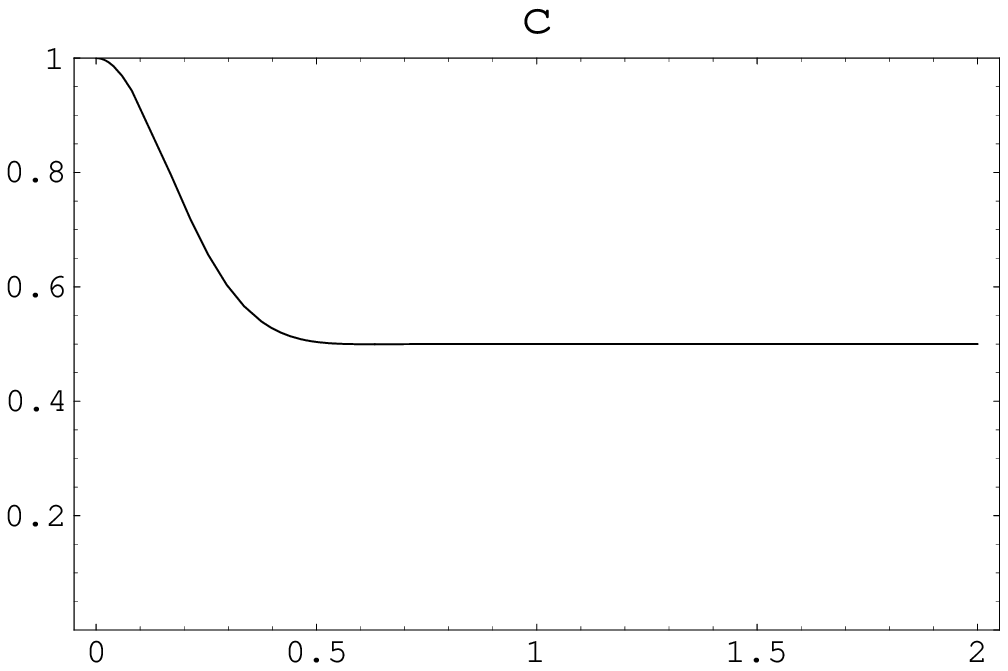}
\> \epsfxsize=8.3cm\epsffile{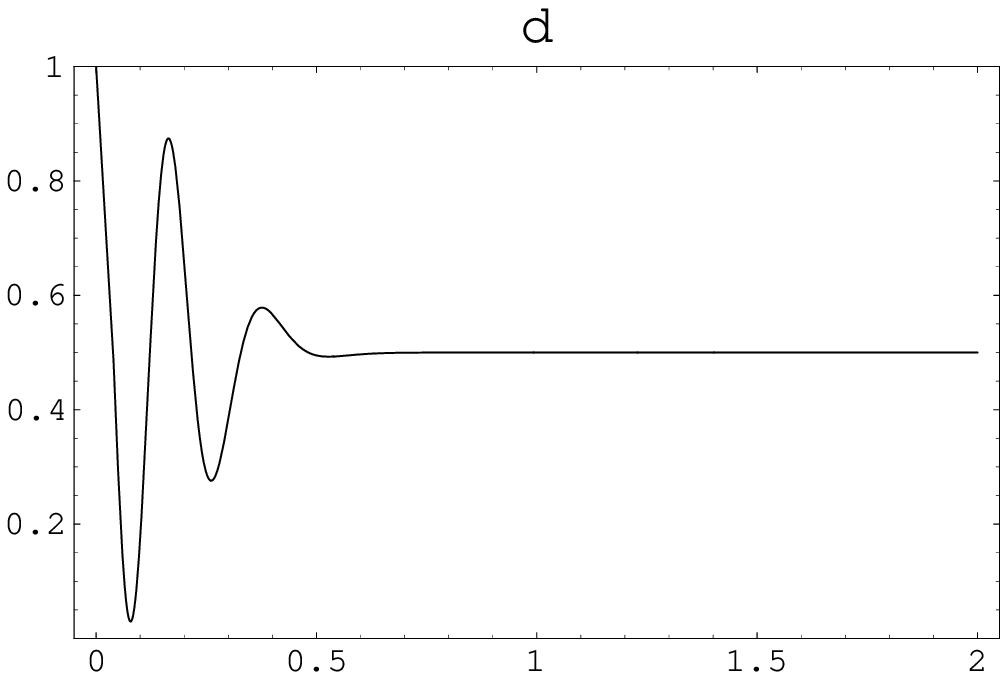}\\
\> \epsfxsize=8.3cm\epsffile{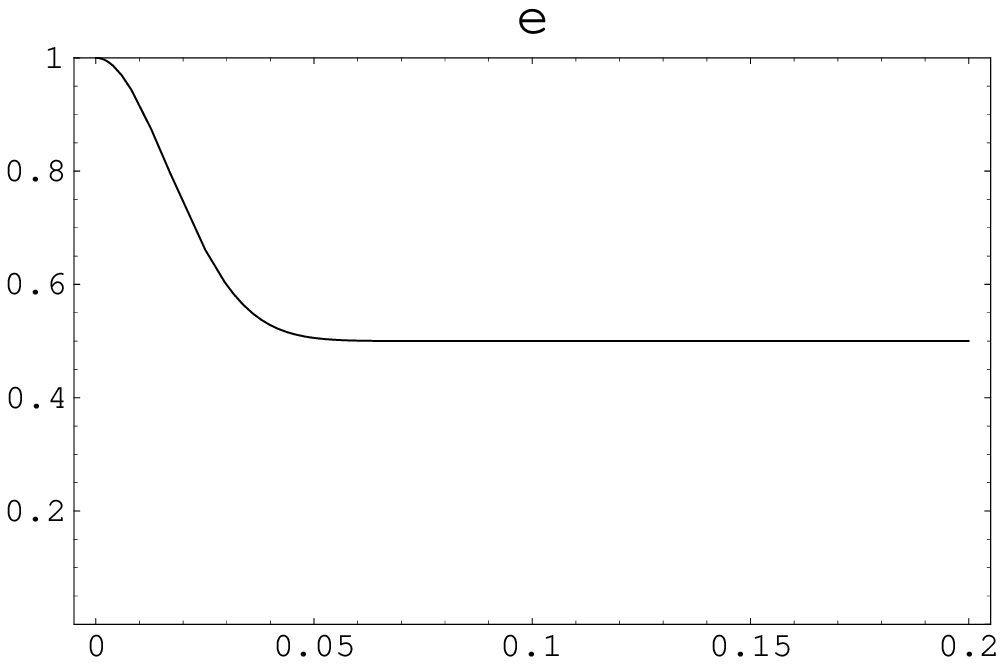} \>
\epsfxsize=8.3cm\epsffile{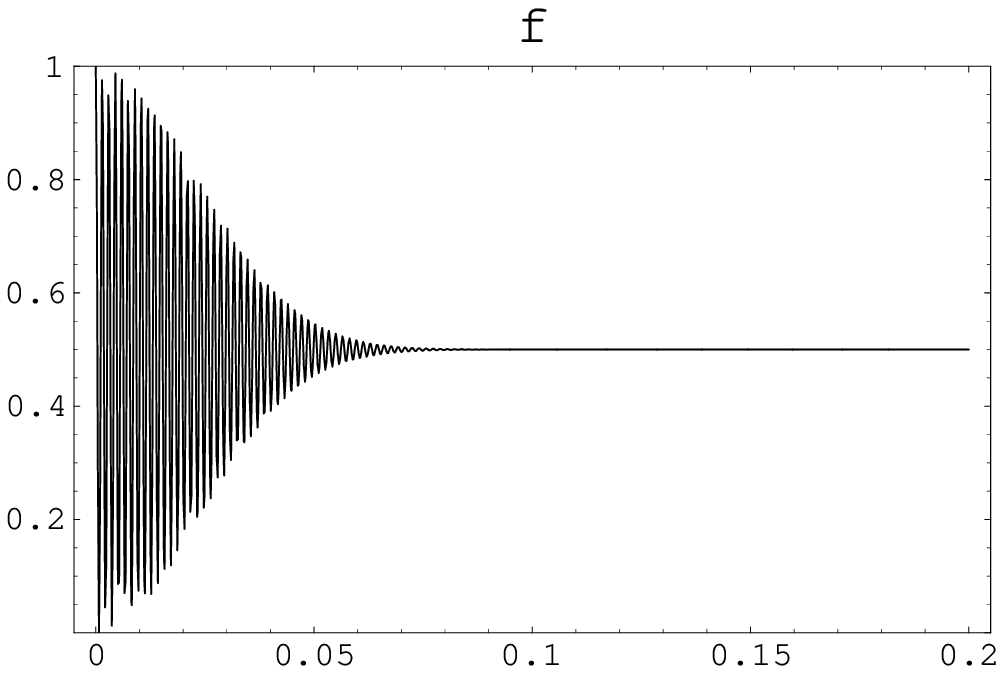}
\end{tabbing}
\caption{The horizontal axe
denotes time period $t^\prime-t$, the vertical
axe denotes the decoherence factor $G[t,t^\prime,\hat{\rho}(0)]$, parameters $%
{\omega_1=0.2,\omega_2=1.3,d_e=0.8,d_e=0.2,\omega_e=1.0}$, (a)${N=10, t=0}$,
(b)${N=10, t=10}$, (c)${N=10^2, t=0}$, (d)${N=10^2, t=10}$, (e)${N=10^4, t=0}
$,(f)${N=10^4, t=10}$.}
\end{figure}

In FIG.$2$, we observed the similar phenomena as in FIG.$1$. We
would like to emphasize that difference between the number state
and the coherent state only manifests in the case that the number
is quite small. We can expect that they may gives the same limit
for infinite particle number. Indeed, the results are identified
by our numerical simulation.

\section{Concluding Remark}

In fact, embodying the wave nature of particles in the quantum
world, the quantum coherence is usually  reflected by the spatial
interference of two or more ``paths'' in terms of single particle
wave function. However, the usual quantum coherence phenomenon
with the first order interference fringes does not sound very
marvellous for the same circumstances can also occur in classical
case, such as an usual optical interference. But in association
with the Hanburg-Brown-Twiss experiment [7], Glauber's higher
order quantum coherence manifests the intrinsically quantum
features of coherence beyond the classical analogue. For example,
in a quantum system composed by identical particles, the quantum
coherence is indeed manifested in the observation of interference
fringes reflected  not only by the first order correlation
functions, but also by higher order ones .

On the other hand, in the present of external quantum system (i.e.
an apparatus) interacting with the studied system, the quantum
decohernce of the system happens as the disappearance of the first
order interference fringes. This decoherence mechanism provides
the essential elements in the understanding for quantum
measurements and the transition from quantum to classical
mechanics. Just based on this conception, our present work extends
the above understanding for quantum decoherence in terms of the
first order interference to the high order case. With a two mode
boson model, we have studied the second order decoherence in the
classical limit. Even without the factorization structure and thus
the obvious the macroscopic limit, the high-order quantum
decoherence still happens in the classical limit, i.e., when the
quantum number to infinity. It is concluded that this decoherence
process losing the higher order coherence can be also explained as
a generalized ``which-path'' measurement for the defined
multi-particle paths.

\vskip 0.2cm {\bf Acknowledgement} This work is supported by NSF
of China and the knowledged Innovation Programme(KIP) of the
Chinese Academy of Science.
%\end{multicols}
%\begin{multicols}{2}

%\end{thebibliography}

\end{document}